\documentclass[10pt,aps,prd,twocolumn,showpacs,superscriptaddress,nofootinbib,nobibnotes,longbibliography,floatfix]{revtex4-2}

\usepackage[utf8]{inputenc}
\usepackage[T1]{fontenc}
\usepackage{bm}
\usepackage{mathtools,amsmath,amssymb,amsfonts,mathrsfs,eucal,graphicx,tensor,csquotes,accents,commath,chngcntr,siunitx}
\usepackage[dvipsnames]{xcolor}
\usepackage[unicode]{hyperref}
\hypersetup{colorlinks=true, citecolor=MidnightBlue,
            linkcolor=MidnightBlue, urlcolor=MidnightBlue, linktocpage=true}
\usepackage[normalem]{ulem}
\usepackage{orcidlink}

\newcommand{\orcid}[1]{\href{https://orcid.org/#1}{\includegraphics[width=10pt]{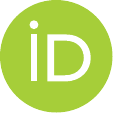}}}

\begin{document}

\author{Saulo Albuquerque \orcid{0000-0003-2911-9358}}
\affiliation{Departamento de F\'isica, Universidade Federal da Para\'iba, Caixa Postal 5008, Jo\~ao Pessoa 58059-900, PB, Brazil}
\affiliation{Theoretical Astrophysics, IAAT, University of T\"ubingen, D-72076 T\"ubingen, Germany}
\email{saulo.filho@academico.ufpb.br}

\author{Sebastian H. V\"olkel \orcid{0000-0002-9432-7690}}
\affiliation{Max Planck Institute for Gravitational Physics (Albert Einstein Institute), D-14476 Potsdam, Germany}
\email{sebastian.voelkel@aei.mpg.de}

\author{Kostas D. Kokkotas \orcid{0000-0001-6048-2919}}
\affiliation{Theoretical Astrophysics, IAAT, University of T\"ubingen, D-72076 T\"ubingen, Germany}

\author{Valdir B. Bezerra$^1$ \orcid{0000-0001-7893-0265}}

\date{\today}

\title{Inverse problem of analog gravity systems II: Rotation and energy-dependent boundary conditions}

\begin{abstract}
In this work, we study the inverse problem of analog gravity systems which admit rotation and energy-dependent boundary conditions. 
By extending two recent results, we provide a recipe that allows one to relate resonant transmission spectra with
effective potentials and even reconstruct the boundary condition at the core. 
Our methodology is based on the WKB method and the identification of universal features in the transmission. 
One of the main advantages of this method is that it is parameter free, and  relies only on general properties of the underlying potential, instead of specific models. 
While the reconstruction of underlying potentials is generally not uniquely possible, the inverse method provides effective potentials with similar spectral properties to the original one.    
To demonstrate the accuracy and scope of our method, we apply it to a rotating imperfect draining vortex, which has been proposed as an analog system to astrophysical extreme compact objects. 
We conclude that the capability to explore energy-dependent boundary conditions could be of interest for experimental studies of such systems.  
\end{abstract}

\maketitle

\section{Introduction}\label{intro}

The pioneering idea of probing astrophysical gravitating systems with lab-controlled analog experiments \cite{Unruh:1980cg} has received much attention and significance in the past 20 years. 
New experimental setups and analog physical systems have been proposed together with analytical solutions \cite{bec1,bec2,hawkingbec1,hawkingbec2,hawkingbec3,Novello2002,Volovik2003, Unruh2007, Vieira:2021xqw, Steinhauer:2014dra,Steinhauer:2015saa,2022PhRvD.105d5015V,kostashoracio,Solidoro:2024yxi,DelPorro:2024tuw}
, extending the horizons for state-of-the-art research within this investigation line. 
In particular, much experimental interest has recently been raised for analog systems that can be used, in principle, to better understand rotating astrophysical compact objects, mainly black holes. 
Experiments conducted with acoustic analogs to black holes have detected, for example, the analog of Hawking radiation \cite{Rousseaux:2007is, PhysRevLett.117.121301, 2019Natur.569..688M, 2021NatPh..17..362K}, propagation of light in optical fibres \cite{2010PhRvL.105t3901B}, laser pulse filaments  \cite{Philbin:2007ji}, and Bose–Einstein condensates \cite{Lahav:2009wx}. In addition, superradiant effects for the scattering profile of acoustic incident waves have been studied \cite{PhysRevLett.117.271101,2017NatPh..13..833T}, 
as widely known, superradiance is intrinsically related to black holes' rotation \cite{zeldovich1971generation,zeldovich1972amplification}. 
Analogs of black and white hole horizons in superfluid $^3\text{He}-B$ have been created \cite{PhysRevLett.123.161302}, while proposals for using analog simulations of quantum gravity with fluids have been recently proposed \cite{2023NatRP...5..612B}. Furthermore, in another recent experiment, the specific signature of curvature effects from rotating black hole spacetimes was measured by studying a giant rotating vortex in superfluid ${^4}$He \cite{patrik}.

Among the most important advantages of conducting research with analog systems is the availability of accurate data on their cross sections and scattering properties. 
The physical identification between the evolution of acoustic perturbations in analog systems and the propagation of field perturbations around a black hole is then given by the mathematical equivalence between the master equations describing them \cite{1993gr.qc....11028V,visser2}. 
Since those equations are in the form of a Schr\"odinger-like wave equation, it motivates the application and extension of Wentzel–Kramers–Brillouin (WKB) based inverse problem methods. 

In the literature, the inversion of WKB formulas provides a powerful tool for reconstructing effective scattering potentials in the scattering and bound state problems \cite{lieb2015studies,MR985100,1980AmJPh..48..432L,2006AmJPh..74..638G}. 
It is semianalytic and offers the advantage of a parameter-free and model-agnostic inverse method based solely on the applicability of the direct WKB method. 
This aspect provides an important advantage when one compares it with model-dependent Bayesian techniques. 
This inverse semianalytic method has been extended for the perturbation theory of wormholes \cite{Volkel:2018hwb}, constant-density stars \cite{Volkel:2017ofl,Volkel:2017kfj}, and other exotic compact objects \cite{Volkel:2019ahb,Volkel:2019gpq}. 

This work is an extension of a previous study \cite{Albuquerque:2023lzw}, in which we have demonstrated the applicability of the semianalytic formulas to analog gravity systems. 
We considered, for illustration, the application of the inverse method for the analog model of exotic compact objects consisting of an imperfect draining vortex in a bathtub \cite{Torres:2022bto}. 
Our main result is the reconstruction of the effective potential associated with the effective background geometry. 
Moreover, we could also infer the reflectivity boundary parameter when it was an energy-independent constant.

Our application in Ref.~\cite{Albuquerque:2023lzw} was restricted to nonrotating systems because rotating systems admit an energy dependence for the scattering potential. 
In such cases, distinct potentials are obtained for different energies of the incident waves. 
In Ref.~\cite{Albuquerque:2024xol}, we have applied the WKB techniques to energy-dependent potentials with two classical turning points. 
Our method proposes an energy-independent, effective potential that reproduces the desired physical properties of the original family of energy-dependent ones. 
We called this reconstructed potential ``WKB-equivalent'' potential because it shares the same bound states (or transmission) we used for its reconstruction. 

Finally, in this work, we extend the inverse method for energy-dependent scattering potentials describing rotating analog gravity systems and for energy-dependent reflectivity parameters defined at their internal boundaries. 
As an application, we extend our previous work~\cite{Albuquerque:2023lzw} by studying the imperfect draining vortex with a nonzero rotation coefficient. 
We can successfully reconstruct effective potentials admitting similar spectral properties as the original ones. Moreover, we can robustly infer the energy dependence of boundary conditions, which could potentially represent internal states of the core. 
For instance, if it admits internal modes that could be excited, one would expect that frequencies close to resonances would be absorbed or emitted. 
In the context of black hole area quantization~\cite{Bekenstein:1974jk,Mukhanov:1986me,Bekenstein:1995ju}, a frequency-dependent reflectivity has been discussed in Refs.~\cite{Hod:2015qfc,Foit:2016uxn,Cardoso:2019apo,Coates:2019bun,Laghi:2020rgl}. 
Our approach thus provides a novel tool for exploring the experimental data measured by suitable types of analog gravity experiments. 

This paper is organized as follows. 
First, we review the numerical setup for computing the direct transmissions used as input for our inverse method in Sec.~\ref{meth1}. 
Second, we present the WKB formulas and their inversions in Sec.~\ref{meth2}. 
Then, in Sec.~\ref{meth3}, we discuss how one can tackle the reconstruction of the reflectivity parameter at the boundary condition in one of the extremities. 
We focus our investigation on the rotating imperfect draining vortex in the application Sec.~\ref{app_results}. 
In Sec.~\ref{res1} and Sec.~\ref{res2}, we compute the effective WKB-equivalent potentials and their associated transmission coefficients. 
In Sec.~\ref{res3}, we test the effective techniques for reconstructing the core reflectivity parameter. 
Finally, in Sec.~\ref{conclusions}, we finish this paper with the conclusions and final remarks.

\section{Methods}\label{methods}

\begin{figure}
\centering
\includegraphics[width=1.0\linewidth]{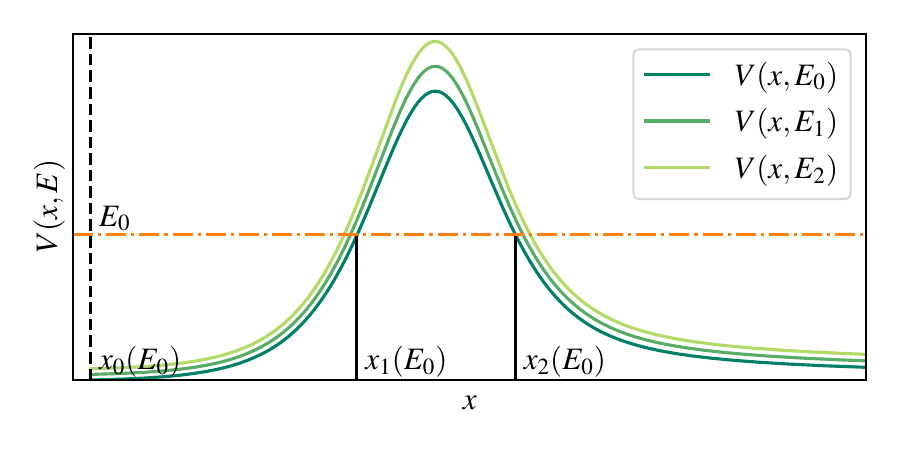}
\caption{Here we qualitatively sketch an energy-dependent potential barrier $V(x,E)$ with three turning points $x_0(E), x_1(E), x_2(E)$ and a reflective surface (dashed vertical line). Since there is a family of potentials depending on $E$, we show $V(x,E)$ for selected values $E_0,E_1,E_2$ (different colors). The turning points are shown only for $E=E_0$.}
\label{fig1}
\end{figure}

Throughout this work, we consider systems that can be modeled by the one-dimensional wave equation
\begin{align}\label{wave_eq}
\frac{\text{d}^2}{\text{d}x^2}\psi(x) + \left[E-V(x,E) \right]\psi(x)=0.
\end{align}
Here $V(x,E)$ is, in general, an energy-dependent potential that describes the system under investigation. 
It is characterized by three turning points defined by $V(x, E)=E$, as shown in Fig.~\ref{fig1}. 
The energy dependence of the potential introduces nontrivial complications in the study of the inverse problem, which is the main objective of this work. 
We review the direct problem in Sec.~\ref{meth1} and then discuss our methods for the inverse problem in Secs.~\ref{meth2} and \ref{meth3}. 
Because of different conventions in the literature, we also define $\omega^2 \equiv E$. 

\subsection{Direct problem}\label{meth1}

In the following, we first outline our numerical method to compute transmission curves in Sec.~\ref{meth1_num}, and then how WKB theory can be used to understand properties of the related resonances in Sec.~\ref{meth1_wkb}.

\subsubsection{Numerical integration}\label{meth1_num}

We briefly summarize the numerical approach used to integrate Eq.~\eqref{wave_eq} and solve the direct scattering problem to obtain the transmission spectrum. 
For more details, we refer the interested reader to Refs.~\cite{Torres:2022bto,Albuquerque:2023lzw,Albuquerque:2024xol}. 

The boundary conditions for scattered waves outside the rotating vortex are given by
\begin{align}\label{BCs1}
\psi(x \approx x_0) \sim&  A^\text{wall} \left[e^{- \mathrm{i} \tilde{\omega}x} 
+ K e^{- 2 \mathrm{i} \tilde{\omega}x_0 }  e^{ \mathrm{i} \tilde{\omega}x} \right], \\
\psi(x \rightarrow \infty) \sim& A^\text{in} e^{-\mathrm{i} \omega x} +A^\text{out} e^{+\mathrm{i} \omega x},
\end{align}
where $\tilde{\omega}$ can in general be different from $\omega$\footnote{In our application in Sec.~\ref{app_results} it is defined as $\tilde{\omega} \equiv \omega - m C$.}. 
$K$ is the reflectivity parameter at $x_{0}$ and it can in general depend on $\omega$. 
One can now define the reflection and transmission coefficients from the amplitudes $(A^\text{in},A^\text{out})$ as follows
\begin{align}\label{scattering}
|t|^2&= \frac{|A^\text{wall}|^2}{|A^\text{in}|^2}(1-|K|^2)\,, \\  |r|^2&= \frac{|A^\text{out}|^2}{|A^\text{in}|^2}.\label{scattering2}
\end{align}

The general wave solution in the frequency domain can be decomposed into the basis of solutions $(u_{h},u_{\infty})$. 
These linear independent wave solutions are defined in terms of their asymptotic behavior
\begin{align}
u_{h}\sim \biggl\{\begin{array}{ll} e^{- \mathrm{i} \tilde{\omega}x},&   x\rightarrow - \infty, 
\\
A^{-}_{\infty}e^{- \mathrm{i} \omega x}+ A^{+}_{\infty}e^{ \mathrm{i} \omega x}, &      
 x\rightarrow + \infty,\end{array}
\end{align}
and
\begin{align}
u_{\infty}\sim \biggl\{\begin{array}{ll} A^{-}_{h}e^{- \mathrm{i} \tilde{\omega} x}+ A^{+}_{h} e^{ \mathrm{i} \tilde{\omega} x},  &  x\rightarrow - \infty,
\\
e^{ \mathrm{i} \omega x},&   x\rightarrow + \infty.  \end{array}
\end{align}

Expressing the general solution $\psi$ as a linear superposition of these solutions ($u_{h},u_{\infty}$) one obtains the following transformation relations between their coefficients
\begin{align}\label{coef1}
A^{+}_{h}&=\frac{\omega}{\tilde{\omega}}A^{-}_{\infty}, \\
\frac{A^\text{in}}{A^\text{wall}}&=\frac{\tilde{\omega}}{\omega}\left(A^{+}_{h}-A^{-}_{h}K e^{- 2 \mathrm{i} \tilde{\omega}x_0 }\right), \label{coef2} \\
A^\text{wall}&=\frac{\omega}{\tilde{\omega}}\frac{\left( A^{-}_{\infty}A^\text{out}-A^{+}_{\infty}A^\text{in} \right)}{K e^{- 2 \mathrm{i} \tilde{\omega}x_0  }} \label{coef3}.
\end{align}
These fundamental relations are needed for computing the scattering coefficients in Eqs.~\eqref{scattering} and \eqref{scattering2}.

\subsubsection{WKB methods}\label{meth1_wkb}

In the following, we outline some basic WKB methods to gain a more intuitive understanding of the complex structure of resonance curves in Sec.~\ref{app_results} and motivate the inverse problem; see Ref.~\cite{bender1999advanced} for a standard textbook on WKB theory. 

Bound states in a potential well can be conveniently computed by the Bohr-Sommerfeld (BS) rule
\begin{align}\label{cBS}
\int_{x_0}^{x_1} \sqrt{E_{n}-V(x,E)} \text{d}x = \pi \left(n+ \phi \right),
\end{align}
where $(x_{0},x_{1})$ are the turning points of the cavity at the energy level $E_{n}$. 
The phase $\phi$ depends on the matching conditions of the cavity, \textit{e.g.}, it is $\phi=1/2$ for a potential well with two exponentially decaying solutions past the turning points, and $\phi=1/4$ if it is set to zero on one side. 
For the rest of this work, we set $\phi=1/4$. Equation~\eqref{cBS} can also be used to approximate the real part of quasistationary states $E_n = E_{0n} + \mathrm{i} E_{1n}$, which are related to resonances in the transmission. 
The imaginary part of quasistationary states can be computed with the Gamow formula
\begin{align}\label{gamow_formula}
\begin{split}
E_{1n} = - &\frac{1}{2}
\left(T_1(E_{0n}) + T_2(E_{0n})\right)
\\
&\left(\int_{x_0}^{x_1}\frac{1}{\sqrt{E_{0n}-V(x,E_{0n})}} \text{d}x \right)^{-1},
\end{split}
\end{align}
where in this work, $T_1(E)$ is the transmission through the reflective boundary surface at $x_{0}$, with a model-dependent reflectivity parameter $K$
\begin{align}\label{Kgamow}
    T_{1}(E)=1-K^2.
\end{align}
$T_2(E)$ is the transmission through the potential barrier itself (without considering the subsequent reflective boundary surface). 
It can be approximated by
\begin{align}
T_2(E) &= \exp\left(2 i \int_{x_{1}}^{x_{2}} \sqrt{E_n-V(x,E)} \text{d}x\right),
\label{transmission}
\end{align}
where $(x_{1},x_{2})$ are the energy-dependent turning points of the potential barrier at the energy level $E$.

\subsection{Inverse WKB methods}\label{meth2}

The WKB methods for the direct problem have been ``inverted'' to study the inverse problem for different types of energy-independent potentials with two turning points in Refs.~\cite{lieb2015studies,MR985100,1980AmJPh..48..432L,2006AmJPh..74..638G}, and have been extended to be applicable to one, three, and four turning point potentials in Refs.~\cite{Volkel:2017kfj,Volkel:2018hwb,Volkel:2019gpq,Volkel:2019ahb,Albuquerque:2023lzw}. 
The application of the inverse methods is as follows. 
Given some spectrum of bound states of a potential well, one can reconstruct the separation of turning points, also called the width of the potential, from the integral
\begin{align}\label{Excursion}
L_1(E) = x_1(E)-x_0(E) = \frac{\partial }{\partial E} I(E),
\end{align}
where $I(E)$ is the so-called inclusion and is given by
\begin{align}\label{Inclusion}
I(E) = 2 \int_{E_\text{min}}^{E}\frac{n(E^\prime)+1/4}{\sqrt{E-E^\prime}} \mathrm{d}E^\prime.
\end{align}
Here $E_\text{min}$ is the minimum of the potential defined by extrapolating where $n(E)+1/4=0$. 
In the case of a given transmission through a two-turning point potential barrier, the separation of turning points is given by 
\begin{align}\label{widthbarrier}
L_2(E) &= x_1(E)-x_0(E) 
\\
&=  \frac{1}{\pi} \int_{E}^{E_\text{max}} \frac{\left(\text{d}T(E^\prime)/\text{d}E^\prime \right)}{T(E^\prime) \sqrt{E^\prime -E}} \text{d} E^\prime.
\end{align}
The key result is that the reconstruction of the potential is not unique but only described by the family of potentials sharing the same separation of turning points. 
Within the validity of the WKB method, this implies they can be ``shifted'' and ``tilted'' without changing their spectral properties, and are thus also called width-equivalent potentials; see Refs.~\cite{Bonatsos:1992qq,Volkel:2018czg}. 

In Ref.~\cite{Albuquerque:2024xol}, the inverse WKB methods had been applied to bound states and transmission curves of energy-dependent potential wells and barriers with two turning points. 
Because of the energy dependence, the structure of the potential is more involved. 
Although one can still define width-equivalent potentials, they are, in general, not isospectral anymore. 
However, energy-independent potentials constructed from the inverse methods can, in principle, still be isospectral to the energy-dependent ones. 
This was confirmed by computing the spectral properties of the WKB inverse potentials with numerical methods as in Sec.~\ref{meth1_num}.

\begin{figure}
\centering
\includegraphics[width=1.0\linewidth]{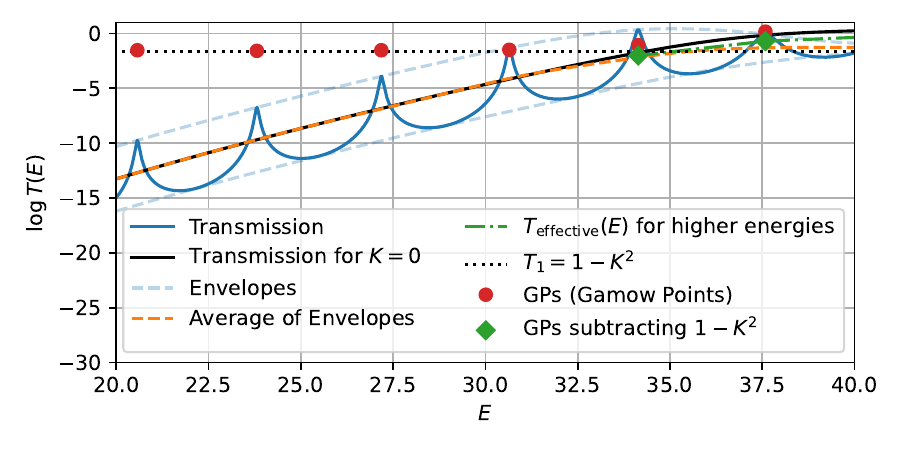}
\caption{The application of the average-envelope method for obtaining the effective transmission $T_\mathrm{effective}(E)$ (red dot-dashed line) approaching the effective potential barrier's transmission (black solid line). Notice that around $E_\mathrm{vertex}$ the average-envelopes curve plateaus at the transmission through the reflective barrier $1-K^2$. For this high-energy regime, the new Gamow points (green dots), obtained by subtracting $1-K^2$ from the original Gamow points (red dots), give us a good approximation for the remaining barrier transmission. For this scenario, $K$ was assumed to be constant.
}
\label{average_envelope}
\end{figure}

In this work, we follow the procedure of our previous work~\cite{Albuquerque:2023lzw} which starts from a transmission curve with resonances. 
The main steps are to identify the location of peaks and their widths to obtain the quasistationary states of the cavity, and to construct an average of two envelopes that can capture the direct transmission through the effective potential barriers. We illustrate the calculation of the latter in Fig.~\ref{average_envelope}. Finally, for computing $L_2(E)$ from this effective barrier transmission $T_\mathrm{effective}(E)$, we assume $E_\text{max}=E_\mathrm{vertex}$ in Eq.~\eqref{widthbarrier} as proposed in Sec. II C of Ref.~\cite{Albuquerque:2023lzw}.

Then, we merge the reconstructed cavity of width $L_{1}(E)$ with the effective barrier, whose width is $L_{2}(E)$. Finally, we obtain the effective WKB-equivalent potentials by setting $x_0$ as a constant. In this way, we complete the reconstruction of the effective potentials. 

\subsection{Reconstruction of boundary condition}\label{meth3}

In the following, we discuss two complementary approaches for reconstructing the boundary condition described by an energy-dependent reflectivity $K(E)$, as defined in Eq.~\eqref{BCs1}. 

The first method is called the ``envelope method''. 
It is related to the method used for inferring the barrier transmission (see Fig.~\ref{average_envelope}). 
In the logarithmic scaling, the transmission curves, as well as their upper and lower envelopes, can be normalized by the effective transmission through the barrier, or similarly, by the average of the envelopes when the last converges to $T^{\text{effective}}_{K=0}$. 
The normalized total transmission is a quasiperiodic oscillation with narrow peaks at the resonant frequencies and local minima between them. 
The values of the peaks and minima are symmetrical and uniform for energies below $E_{\text{max}}$, as shown in Fig.~3 in Ref.~\cite{Albuquerque:2023lzw}. Their amplitudes are directly related to the reflectivity parameter $K(E)$. 
The analytical formula relating $K(E)$ and this amplitude of oscillation for the normalized transmission is given by 
\begin{equation}\label{normalizedformula}
    \log(T_{\pm}^{\text{normalized}}) = \pm\log\left(\frac{1+K(E)}{1-K(E)}\right),
\end{equation}
for the upper $(+)$ and lower $(-)$ envelopes. 
Therefore, if we normalize the logarithmic transmission by the average of its envelope, the reflectivity function can be computed by solving the difference between these envelopes for the reflectivity parameter $K(E)$ in Eq.~\eqref{normalizedformula}. 

Since the envelopes are constructed by interpolating the local maxima or minima, respectively, there is an intrinsic uncertainty related to the interpolation method. 
As long as $K(E)$ is changing less drastically compared to the maxima and minima, the interpolation error can be expected to be small. We quantify this error in Sec.~\ref{res3}.

The second method is called the ``Gamow method'' because it is based on the Gamow formula Eq.~\eqref{gamow_formula} in order to connect the width of the resonance peaks with the transmission at the boundary condition. 
It has previously been explained and applied in Ref.~\cite{Albuquerque:2023lzw}, where more details can be found. 
Because of the underlying assumption required to derive the Gamow formula from a more complex, generalized BS rule for three turning points~\cite{1991PhLA..157..185P}, the transmission through the boundary needs to be very small. 
This implies that $1-K(E)$ needs to be small, and the accuracy of the method is expected to decrease if this condition is violated, which was also observed in Ref.~\cite{Albuquerque:2023lzw}. 

To extend the range of validity also to larger values of $1-K(E)$, we construct the following correction 
\begin{align}\label{gamow_new}
K^\text{new}(E)
&=K^\text{Gamow}(E) + \Delta K(E) 
\\
&\approx 1 + a(1-K(E))  + b(1-K(E))^2 
\end{align}
where $K^\text{Gamow}(E)$ is the leading order earlier result obtained by Eq.~\eqref{Kgamow}, and $a,b$ are constants that depend on the specific case under consideration. 
Note that $K^\text{Gamow}(E)$ is only available at discrete values of $E=E_n$ corresponding to the location of the resonance. 
By identifying 
\begin{align}
K^\text{Gamow}(E)&= 1 + a(1-K(E)), 
\\
\Delta K(E)  &= b(1-K(E))^2,\label{gamow_new2}
\end{align}
one would expect the correction $\Delta K(E)$ to be only depending on the functional form of $K(E)$, not directly on the energy. 
Instead of predicting $a,b$ explicitly, which is nontrivial, we will use the envelope method to compute the correction as
\begin{align}\label{correctinggamowwithenvelope}
\Delta K(E) = K^\text{envelope}(E) - K^\text{Gamow}(E),
\end{align}
and then numerically fit the constant $b$ via Eq.~\eqref{gamow_new2} by assuming $b$ does not depend on $E$, and is thus the same constant for all discrete $K^\text{Gamow}(E)$.

One remark is needed. 
At first, it may seem circular to use one method to compute the necessary correction of another one. 
However, in cases where the envelope method is applied to rapidly changing $K(E)$, the interpolation error can become significant in such energy ranges but very accurate where $K(E)$ changes slowly. 
The numerical fitting of $b$ can effectively average out such uncertainties, and thus make the ``corrected'' prediction Eq.~\eqref{gamow_new} more robust. 
We illustrate this in Sec.~\ref{res3}.

\section{Application and results}\label{app_results}

\subsection{Imperfect draining vortex model}\label{vortex}

As proposed in Ref.~\cite{Torres:2022bto}, an imperfect draining vortex can be regarded as an analog model of an astrophysical extremely compact object. The evolution of small acoustic perturbations in its surroundings is described by a wave equation, which can be separated into a radial and an angular equation. The radial part can be written as in Eq.~\eqref{wave_eq}, where the effective potential is given by
\begin{align}\label{radialpot}
V(r,E)=&-\frac{(mC)^2}{r^4}+\left(1-\frac{1}{r^2}\right)\left(\frac{m^2-\frac{1}{4}}{r^2}+\frac{5}{4r^4}\right) \nonumber \\& +2\frac{mC\sqrt{E}}{r^2}. 
\end{align}
Here $m$ is an integer number that labels the harmonic decomposition of the angular coordinate (azimuthal number), while $C$ characterizes the tangential velocity profile of the vortex. In Eq.~\eqref{wave_eq}, $x(r)$ is the so-called tortoise coordinate.
Notice that the energy dependency couples to the effective potential through the rotation parameter $C$.

To complete the physical model, we introduce an energy-dependent reflectivity $K(E)$ defined at the boundary surface in $r_0=r_h (1+\epsilon)$, where $r_h$ is the acoustic horizon radius in radial coordinate, and $\epsilon$ is a very small parameter. 
Physically, this reflective boundary condition is defined as a cylindrical interface radially displaced by a small radial distance $\epsilon$ from the horizon surface $r_h$. 
From now on, we assume that $\epsilon=2e^{-20}$, which implies that the tortoise coordinate at this reflective boundary is given by $x_0=x(r_0)=-9$. For more details, we refer the interested reader to Ref.~\cite{Torres:2022bto}.

\subsection{Results for inverse problem}

In the following, we apply the inverse method introduced in Sec.~\ref{meth2} and Sec.~\ref{meth3} to the various transmission curves of the imperfect draining vortex from Sec.~\ref{vortex}. 
In Sec.~\ref{res1}, we analyze the impact of the rotation parameter $C$ by varying it until we reach our method's applicability limit. 
Then, in Sec.~\ref{res2}, we discuss how the value of the harmonic constant $m$ impacts the accuracy of the method. 
Finally, in Sec.~\ref{res3}, we demonstrate the performance of our method to infer the energy-dependent reflectivity parameter $K(E)$ by investigating a set of non-trivial examples. 

\subsubsection{Dependency of rotation profile $C$}\label{res1}

We consider different applications for varying rotational velocity profiles $C$. Our analysis starts from transmission curves numerically obtained for cases with $K=0.9$ and $m=10$. 

\begin{figure}
\centering
\includegraphics[width=1.0\linewidth]{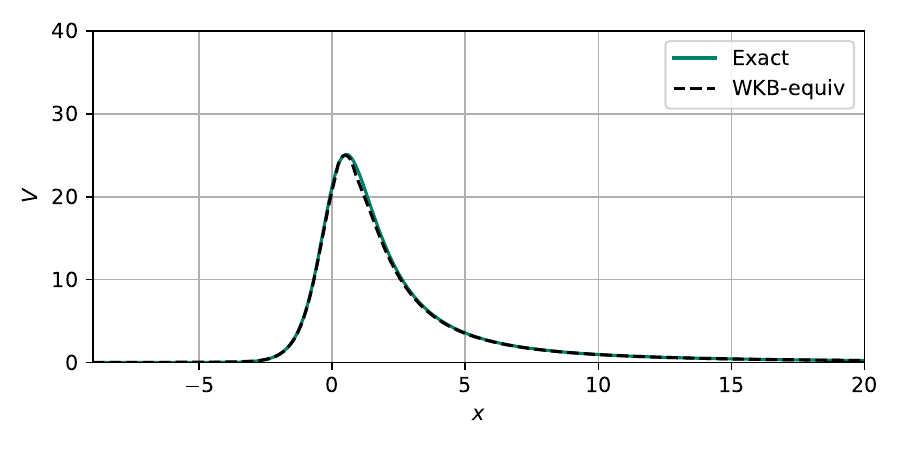}
\includegraphics[width=1.0\linewidth]{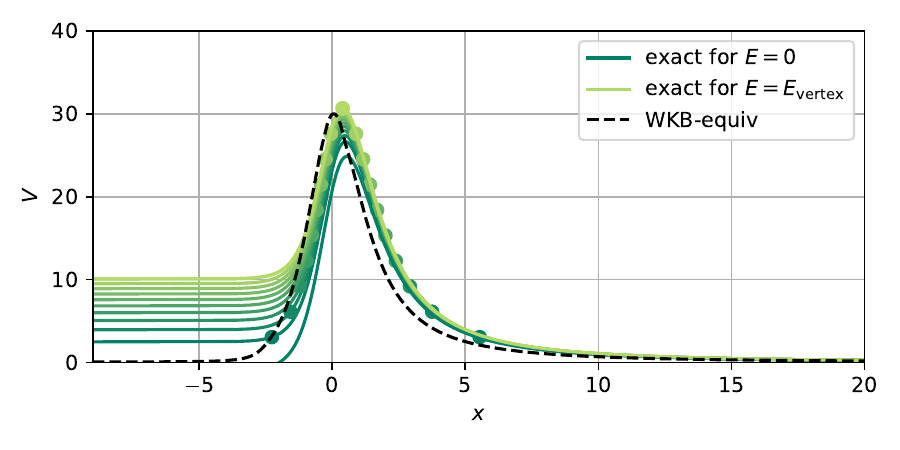}
\includegraphics[width=1.0\linewidth]{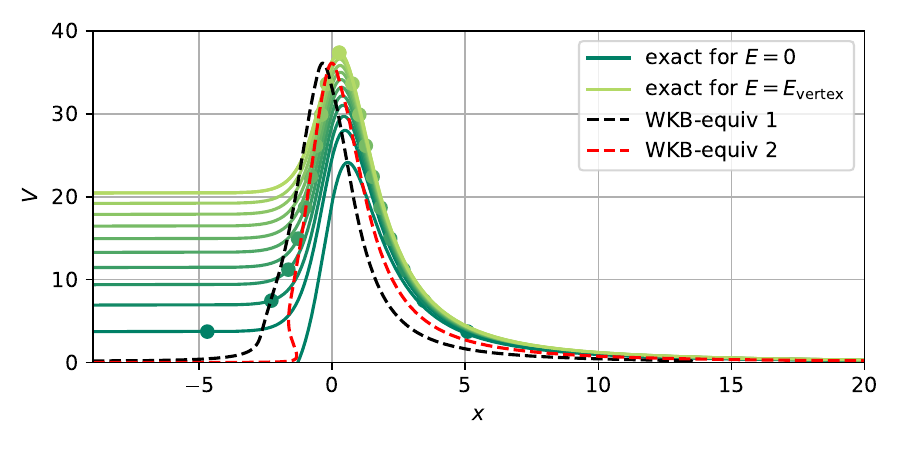}
\caption{In the three panels, we show the WKB-equivalent potential reconstructed by the inverse method presented in this framework for $m=10$ and $C=0, C=0.1$ and $C=0.2$, respectively. 
The solid colored lines are the original energy-dependent potentials from Eq.~\eqref{radialpot} for a range of energies from $E=0$ to $E=E_{\mathrm{vertex}}$. 
A pair of colored dots mark the intersection of each $V(x,E)$ curve with the horizontal line placed at its associated energy value $E$. 
The color scheme is defined by a green-yellow transition from the lowest energy value to the highest. 
In the bottom panel, with $C=0.2$, the red-dashed curve (WKB-equiv $2$) represents the reconstructed potential by taking into account all energy modes. 
The regularized version of the WKB-equivalent potential (WKB-equiv $1$), obtained by dismissing the fundamental mode in the inverse reconstruction, is represented by the black-dashed curve. 
}\label{pannelpotentialsandtransmission}
\end{figure}

We report some examples of those reconstructions in Fig.~\ref{pannelpotentialsandtransmission}, where we compare the effective reconstructed potentials with the corresponding family of energy-dependent potential curves $V(x,E)$ for a certain range of energy values. 
Furthermore, for each energy value, we also present the associated pair of turning points $x_1(E)$ and $x_2(E)$. 

As discussed in Ref.~\cite{Albuquerque:2024xol}, the effective WKB-equivalent potentials and the width-equivalent curves are not supposed to be similar for the method to work. 
While the effective WKB-equivalent potential succeeds in approximately reproducing the physical properties used for its reconstruction, the width equivalent potentials, obtained by continuously connecting the turning points $x_1(E)$ and $x_2(E)$ for each energy $E$, do not provide similar results. 
However, although those curves are increasingly different for increasing values of the rotation parameter $C$, there is still a remarkable agreement between the original value of $E_\mathrm{vertex}$, obtained analytically from Eq.~\eqref{radialpot}, with the $E_\mathrm{vertex}$ value inferred by the turning points' analysis from Sec.~\ref{meth2}.

Notice in Fig.~\ref{pannelpotentialsandtransmission} that, for higher rotations, e.g., $C= 0.2$, there is an ``overhanging cliff'' \cite{lieb2015studies} behavior on the width of the cavity with $E$, which comes from a nonmonotonic dependence of $L_{1}$ with $E$. 
This problematic behavior of $L_{1}$ is explained by the fact that the inverse BS rule Eq.~\eqref{Excursion} is reading a $n$-dependence of $E_{0n}$ that grows faster than $n^2$ for low energies. 
The $n(E_{0n})\propto \sqrt{E_{0n}}$ corresponds to the limiting case of a second fixed wall enclosing the cavity for a confined particle (infinite square box problem in quantum mechanics \cite{GriffithsQM}). 
Since those reconstructed cavities with overhanging cliffs cannot represent physical solutions, we constrain the applicability of our inverse method here for rotation parameters $C$ equal or lower than $C=C_\mathrm{crit}$, where $C_\mathrm{crit}$ is the highest rotation coefficient for which the overhanging cliffs behavior does not arise. 
That defines the range of applicability of this method for the imperfect draining vortex with $m=10$. For other $m$ values, a different range in $C$ needs to be further established.

It is the first few modes that produce the nonmonotonic behavior on the energy dependence of the cavity width. 
Suppose we want to recover the monotonic behavior of the effective reconstructed cavity. In that case, we can try repeating the inverse method, but this time neglecting the first mode into the inverse BS rule.  
Depending on the rotation parameter $C$, we might even need to go further and neglect a couple of first modes instead. 
Then, in this way, we recover the monotonic behavior of the reconstructed cavity width, although we lift our potential minimum $E_\mathrm{min}$.

\begin{figure}
\centering
\includegraphics[width=1.0\linewidth]{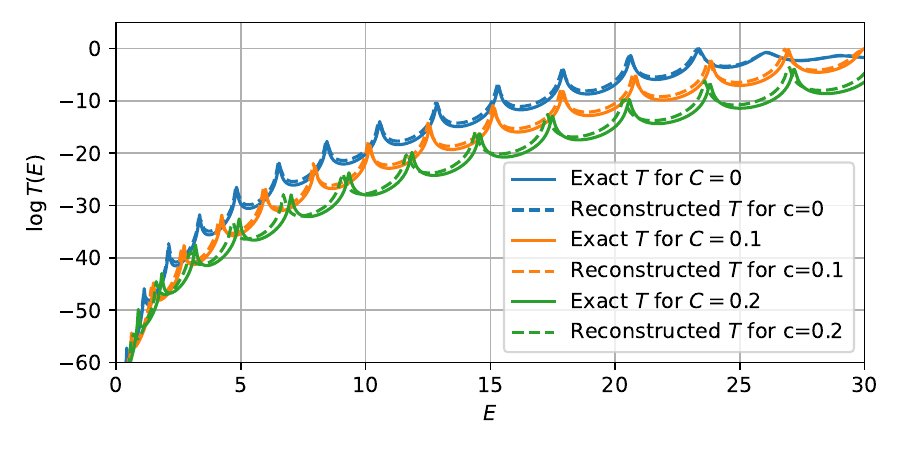}
\caption{
Here we show a comparison between the original transmissions, calculated from the energy-dependent potentials (solid colored line) in Eq.~\eqref{radialpot}, and the reconstructed transmissions calculated from the WKB-equivalent potentials in Fig.~\ref{pannelpotentialsandtransmission} (colored dashed lines). 
}\label{transmm10}
\end{figure}

Finally, we can now test the accuracy of the reconstructed effective potentials to reproduce the transmission curves used as input for their reconstruction and whether it can approximately predict the same resonant peaks from the original curves. 
For this goal, we use the same numerical scheme used for the original energy-dependent potentials \ref{meth1_num}, but now replacing the analytical potential defined in Eq.~\eqref{radialpot} with the effective WKB-equivalent potential. 
The results are shown in Fig.~\ref{transmm10}, where we compare the original transmissions from the energy-dependent potentials (solid lines), with the ones reconstructed from our inverse method (dashed lines). 
Notice the good agreement between the transmission curves and the reproduction of the location and width of the resonant peaks from the original energy-dependent potential for $C=0$ and $C=0.1$.

For rotations above the working limit threshold of our method, such as  $C=0.2 \geq C_\mathrm{crit}$ for instance, the curves with overhanging cliffs forbid us from interpolating the potentials to run the reconstruction of the transmission. 
As discussed previously, however, we can circumvent this problem by neglecting the first mode. In this way we reobtain a functional inverse reconstructed potentials. 
Those potentials have their minimum $E_\mathrm{min}$ vertically lifted upwards, decreasing the accuracy of their reconstruction for energies around the first considered mode. 

Since the reconstruction of higher modes also depends on the lower modes, the inaccuracy does not go away completely for higher energy modes. 
To illustrate that, we present the WKB-equivalent potentials in the bottom panel of Fig.~\ref{pannelpotentialsandtransmission} for $m=10$ and $C=0.2$. 
Here we contrast the reconstructed potential possessing overhanging cliffs (red dashed curve) with its regularized version obtained by inputting all modes but the fundamental one in the inversion formulas (black dashed curve). 
Finally, we also compute the associated reconstructed transmission for this case. 
This result is represented by the green dashed curve in Fig.~\ref{transmm10}. 
We can see the accuracy loss when comparing this reconstructed transmission with the exact one (green solid curve).

\subsubsection{Dependency on the angular parameter $m$} \label{res2}

\begin{figure}
\centering
\includegraphics[width=1.0\linewidth]{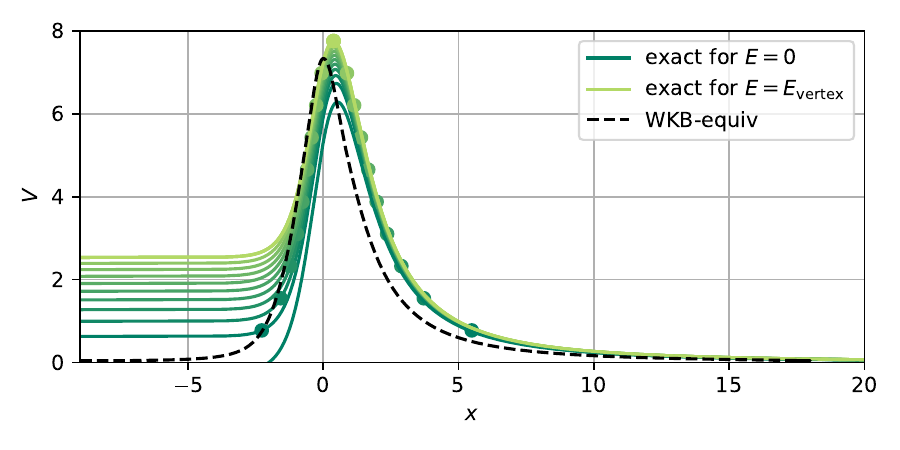}
\includegraphics[width=1.0\linewidth]{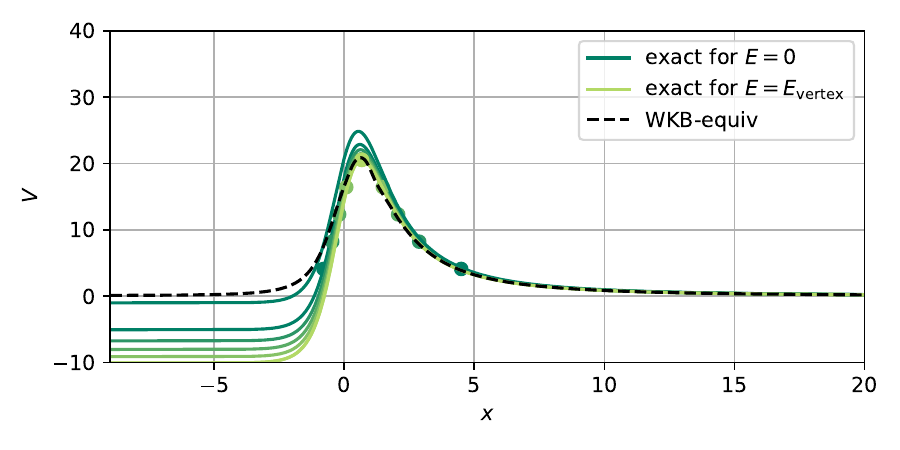}
\includegraphics[width=1.0\linewidth]{Figures/Inverse_Potentials_m10_c0.1_k1_kA0.9.pdf}
\caption{In the three panels, we show the WKB-equivalent potential reconstructed by the inverse method presented in this framework for $C=0.1$ and $m=5, m=-10,~\text{and} m=10$,  respectively. 
They are represented as black dashed curves. The solid colored lines are the original energy-dependent potentials from Eq.~\eqref{radialpot} for a range of energies from $E=0$ to $E=E_{\mathrm{vertex}}$. 
A pair of colored dots mark the intersection of each $V(x,E)$ curve with the horizontal line placed at its associated energy value $E$. 
The color scheme is defined by a green-yellow transition from the lowest energy value to the highest.}\label{wkbequivpotm5}
\end{figure}

\begin{figure}
\centering
\includegraphics[width=1.0\linewidth]{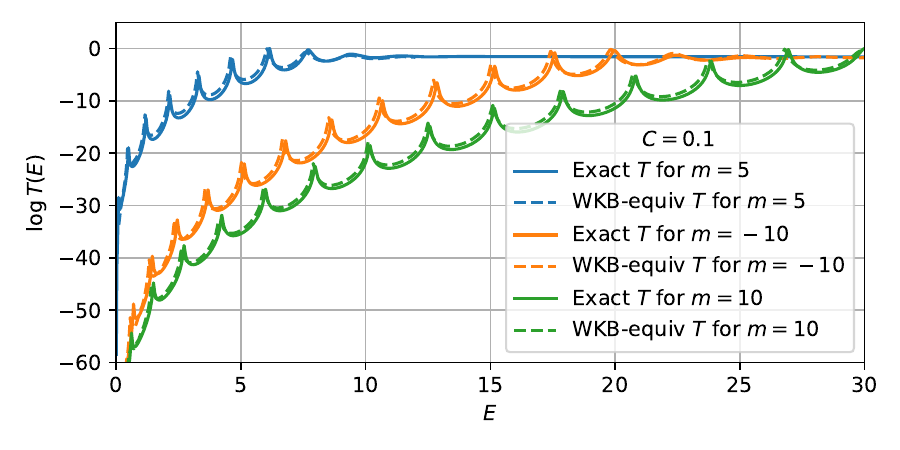}
\caption{Here we show a comparison between the original transmissions, calculated from the energy-dependent potentials (solid colored line) in Eq.~\eqref{radialpot}, and the reconstructed transmissions calculated from the WKB-equivalent potentials in Fig.~\ref{wkbequivpotm5} (colored dashed lines). 
}\label{transmmminus8}
\end{figure}

Now, we investigate the impact of different angular momentum parameters $m$ in our inverse method. In Fig.~\ref{wkbequivpotm5}, we contrast the WKB-equivalent potentials, reconstructed by employing our inverse techniques, and the associated family of energy-dependent potentials for a certain range of energy values. The comparison between the transmissions computed from these energy-dependent potentials and their corresponding WKB-equivalent potentials are shown in Fig.~\ref{transmmminus8}.

As discussed in Ref.~\cite{Albuquerque:2023lzw} for the nonrotating case, $m$ controls the height of the potential barrier. Hence, a larger value of $m$ implies a larger number of resonant quasistationary modes, and accordingly, a higher overall accuracy for the reconstruction. For the more general rotating case, a similar qualitative behavior is expected. We illustrate this in Fig.~\ref{wkbequivpotm5}, where one can see the impact of the absolute value of $m$ for the height of the reconstructed potentials (read $E_\mathrm{vertex}$); and in the associated number of resonant peaks for the transmission curves in Fig.~\ref{transmmminus8}. In terms of the reconstruction's overall quality, we see a considerable difference in the accuracy of the reconstructed transmissions when comparing results for $m=10$ with the ones for $m=5$. This is expected because the underlying WKB theory should become more accurate for large values of $m$.

Furthermore, an additional analysis that can only be carried out when we consider rotating regimes concerns applying our inverse method to investigate the scattering of counterrotating waves, rather than co-rotating ones. For nonzero rotating parameter $C$, the incident acoustic waves can be both oriented in the vortex's rotating/absorbing direction or, oppositely, in a contrary direction. Those scenarios correspond to corotating and counterrotating waves, respectively. Corotating waves are represented by a positive angular momentum $m$, while counterrotating waves have negative $m$. So far, we have only considered the corotating regime. Considering the counterrotating regime is interesting because it can lead to instabilities, especially if large $m$ modes are excited. 

Looking at Eq.~\eqref{radialpot}, we can see that the term coupling the energy dependence to the potential consists of a linear product between $m$ and $C$. In all other terms, $m$ appears squared, so there is no special role played by its sign. The effect of the $m$'s sign only concerns the energy-dependent term, and it corresponds to two aspects; first for $m>0$, increasing the size increment of the potential barriers with $E$ (which is also amplified for larger $|m|$ values); second for $m<0$, ``pulling down'' the potential curves as the energy increases. 
We illustrate this qualitative discussion in the last two panels of Fig.~\ref{wkbequivpotm5}, where we present the results of the inverse method's application for two distinct scenarios for $C=0.1$, namely $m=-10$ (counterrotating regime) and $m=10$ (corotating regime). 

For the first scenario, the amplification of all the potential barrier's heights implies an overall increase of our WKB-equivalent potential's height. Since all $V_\text{max}$ are pushed upwards, it is expected that $E_\mathrm{vertex}$ goes in the same direction. For the second scenario, however, $E_\mathrm{vertex}$ is pushed down just like all the other turning points for any $E$. That behavior opposes the corotating waves. The reconstructed transmissions associated with both scenarios discussed here are depicted in Fig.~\ref{transmmminus8}, where we compare them with their corresponding original transmission.

\subsubsection{Dependency of reflectivity $K$}\label{res3}

\begin{figure}
\centering
\includegraphics[width=1.0\linewidth]{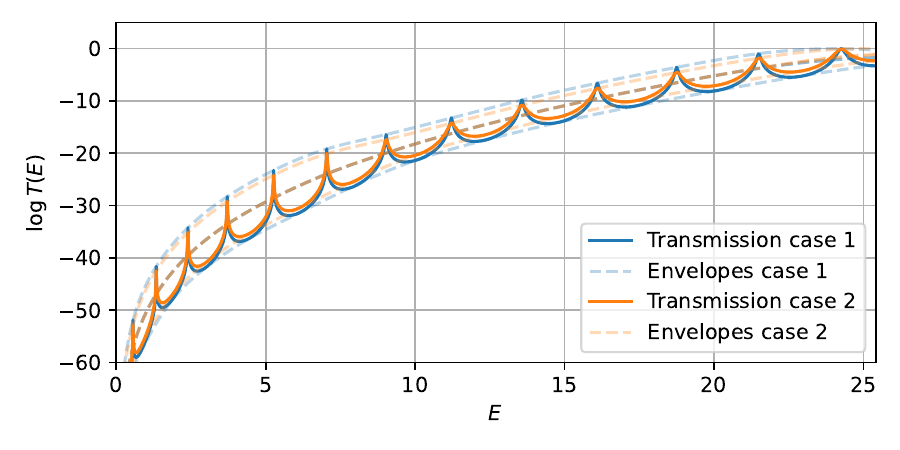}
\includegraphics[width=1.0\linewidth]{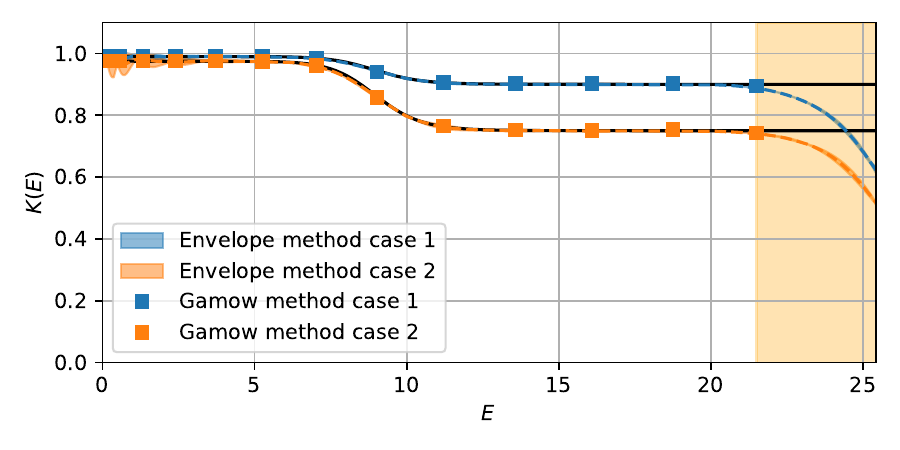}
\caption{In the upper panel we show the transmission for the two reflectivity cases (cases 1 and 2), including their envelopes and the average between them. The injected reflectivity is represented in the lower panel by the solid black line. We also present the final results of our method inference for those two reflectivities by using both the Gamow method (colored squares) and the envelopes' technique (colored dashed lines). These two cases considered here consisted of two plateau-dominated models for the reflectivity, representing transitions from a low-energy plateau to another plateau for higher energies.}\label{transplateaus}
\end{figure}

\begin{figure}
\centering
\includegraphics[width=1.0\linewidth]{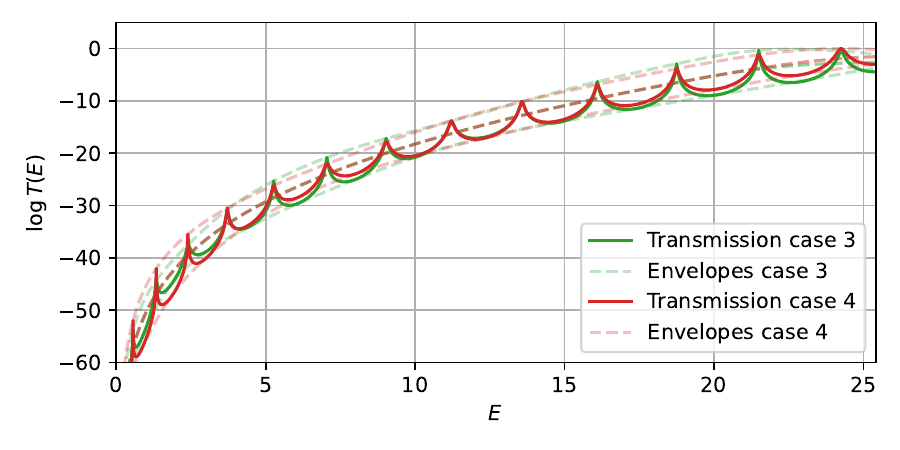}
\includegraphics[width=1.0\linewidth]{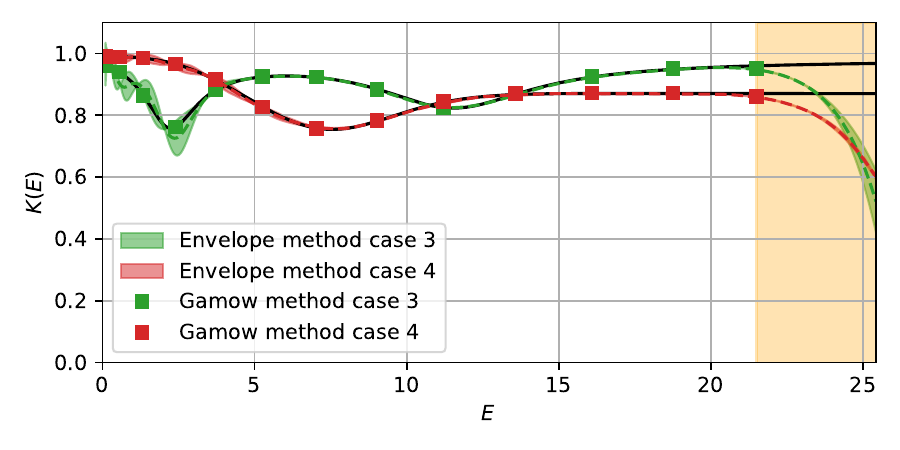}
\caption{In the upper panel, we show the the transmission for the two reflectivity cases (cases 3 and 4), including their envelopes and the average between them. The injected reflectivity is represented in the lower panel by the solid black line. We also present the final results of our method inference for those two reflectivities by using both the Gamow method (colored squares) and the envelopes' technique (colored dashed lines). These two cases considered here consisted of reflectivity models
with some local valleys due to possible core-resonant effects.} \label{transvalleys}
\end{figure}

Finally, we now apply the two approaches introduced in Sec.~\ref{meth3} for inferring the reflectivity parameter at the boundary condition in $x_0$. We expect in advance that both presented methods for the reconstruction of the boundary condition fail around energies close to $E_{\text{max}}=E_{\mathrm{vertex}}$, which defines the range of applicability for our reconstruction. 
In Figs.~\ref{transplateaus} and \ref{transvalleys} we show the results of both approaches for reconstructing some injected reflectivity functions $K(E)$. The expected breakdown of our method is indicated by the orange-shaded area. 
To focus on the properties of $K$ when applying the inverse method, we fix our model to $m=10$ and a rotation profile specified by $C=0.01$. Figure ~\ref{transplateaus} contains reflectivity functions plateauing at two different values of $K$; one for low asymptotic energy and the other one for higher asymptotic energy. 
The physical motivation for this step transition in $K$ comes from the analogy between $K$ with the vortex's overall behavior for the reflectivity coefficient $R$. 
Likewise, we show in Fig.~\ref{transvalleys} how effectively our inference methods tackle reflectivity functions with local valleys in the core-reflectivity. 
Those local valleys are physically motivated by the idea of a scattering core with intrinsic resonant tunneling properties, in analogy with the vortex itself. 
All those injected reflectivities, although physically motivated by general qualitative assumptions, are only models to demonstrate the capabilities of the method. 
It is outside the scope of this framework to derive the core reflectivity functions from a first-principle perspective, as this would depend on the specific physical properties of the example.  

To finish our discussion, we analyze the systematic error associated with the outlined approaches for reconstructing the boundary reflectivity $K(E)$. As we discussed in our earlier analysis in Sec.~\ref{meth3}, the approach based on the envelope's average was originally more reliable and flexible. This technique remained accurate with no regard to the rotation parameter and the absolute value of the reflectivity $K$, as long as $E<E_{\mathrm{vertex}}$. In fact, this method even provides us with the reference values for the reflectivity $K$ that are also used to correct the Gamow method, see Eq.~\eqref{correctinggamowwithenvelope}. As we mentioned in Sec.~\ref{meth3}, we exploit the systematic error introduced by the interpolation to evaluate the systematic errors of this method. Our results are depicted as the colored areas in Figs.~\ref{transplateaus} and ~\ref{transvalleys}.

\begin{figure}
\centering
\includegraphics[width=1.0\linewidth]{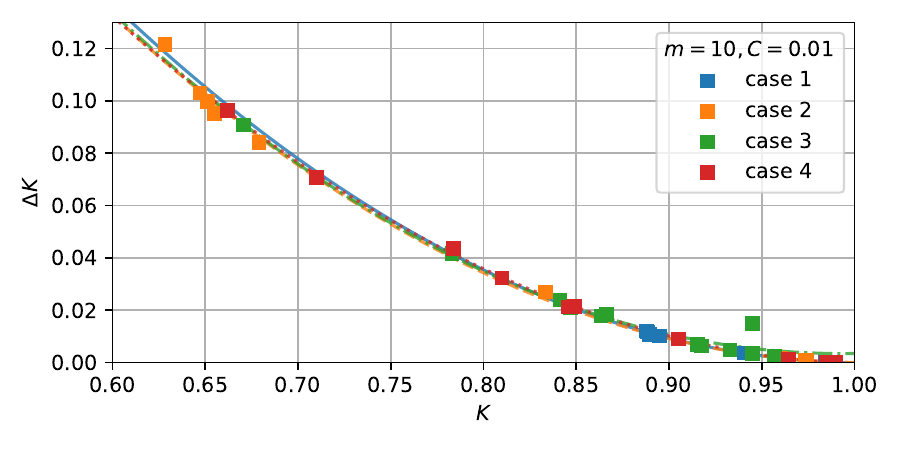}
\caption{Here we show the deviations on the reflectivity $\Delta K(E_n)$ inferred by the Gamow points for the four cases shown in Figs.~\ref{transplateaus} and in \ref{transvalleys}. Those deviations are obtained by the difference between $K^{\text{Gamow}}(E_n)$ and the reflectivity $K^{\text{envelope}}(E)$ predicted by the envelopes approach. For each scenario, the fitting curve for the associated data with the quadratic expansion in Eq~\eqref{gamow_new} is shown. }\label{gamowcorrections}
\end{figure}

Meanwhile, for the Gamow method, we noticed that as we decreased the value of the injected reflectivity, it showed an approximately quadratic deviation from the values obtained by the envelope method. 
This is exactly what we expected from the discussion in Sec.~\ref{meth3}. 
Therefore, the correction proposed by Eq.~\eqref{gamow_new} allows us to circumvent this limitation and refine our findings to obtain the precise reconstruction of the reflectivities we see in Figs.~\ref{transplateaus} and \ref{transvalleys}. 
For each case, we have fitted the deviation data for $\Delta K(E_n)$ with the quadratic correction in Eq.~\eqref{gamow_new2}. 
The resulting fits are represented by the colored curves plotted in Fig.~\ref{gamowcorrections}. 
Notice the approximate convergence between all of those fitting curves. 
This validates the point we made in Sec.~\ref{meth3} that the corrections would not depend on the energy, but only on the absolute value of $K^{\text{Gamow}}$.

\section{Conclusions}\label{conclusions}

In this work, we extended the WKB-based framework presented in Ref.~\cite{Albuquerque:2023lzw} to study the inverse problem of certain types of energy-dependent potentials with energy-dependent boundary conditions. The input for the inverse method is the transmission through the potential. If it admits resonance features, they can be related to quasistationary states of a potential well, and the transmission through a potential barrier and an energy-dependent boundary condition. 

Until recently, the inverse WKB methods have been applied to energy-independent potentials Refs.~\cite{lieb2015studies,MR985100,1980AmJPh..48..432L,Bonatsos:1992qq,2006AmJPh..74..638G,Volkel:2017kfj,Volkel:2018hwb,Volkel:2019gpq,Volkel:2019ahb,Albuquerque:2023lzw}
. From Ref.~\cite{Albuquerque:2024xol} it is known that the construction of isospectral, energy-independent potentials can be possible from bound states and transmissions of energy-dependent ones. This assumes the validity of the WKB approximation in the form of the Bohr-Sommerfeld rule and Gamow formula for two turning-point potentials, respectively.

By applying it to an imperfect draining vortex~\cite{Torres:2022bto}, we explicitly demonstrated how the inverse method can be used to find energy-independent potentials that are isospectral (within the validity of the WKB approximation and for three turning points), which we verified by computing the transmission of the inverse potential numerically. Moreover, our approach also allows the reconstruction of energy-dependent boundary conditions, which we explicitly demonstrated. The latter is highly nontrivial in the inverse problem, because computing its impact on the transmission relies, in general, on the knowledge of the underlying potential. Our method can bypass this limitation by using the reconstructed potential as a placeholder. 

The reconstruction of energy-dependent boundary conditions can be a very interesting tool when applied to experimental measurements of transmissions of applicable analog gravity systems. Here one might not know \textit{a priori} the right boundary conditions, or simply want to verify the experimental setup and underlying assumptions. For example, the boundary condition could effectively describe internal degrees of freedom, e.g., absorption lines of the core. We have demonstrated that the model-independent reconstruction of such absorption features can be possible.   

The methods discussed here are not limited to analog gravity systems. Systems with similar phenomenology have been suggested in a scenario with astrophysical exotic compact objects. Such objects can, in principle, admit quasistationary states and could reveal themselves in terms of so-called gravitational wave echoes after the merger of two such objects. Here the boundary condition is highly speculative and not well understood since different models exist (see Ref.~\cite{Cardoso:2019rvt} for a review of such systems). 

Finally, it has recently been suggested that the high-frequency content of the black hole ringdown can be associated with their respective greybody factors~\cite{Oshita:2023cjz,Okabayashi:2024qbz,Rosato:2024arw,Oshita:2024fzf}. If this identification is robust in measurable frequency ranges, applying the inverse method, i.e., as presented in this work or in Ref.~\cite{Volkel:2019ahb}, could also be possible for gravitational wave observations. The connection of quasinormal modes and greybody factors within WKB theory in this context has recently been discussed in Ref.~\cite{Konoplya:2024lir}.

\acknowledgments
S.~A. acknowledges funding from Conselho Nacional de Desenvolvimento Cient\'ifico e Tecnol\'ogico (CNPQ)-Brazil and Coordena\c{c}\~ao de Aperfei\c{c}oamento de Pessoal de N\'ivel Superior (CAPES)-Brazil. 
S.~H.~V. acknowledges funding from the Deutsche Forschungsgemeinschaft (DFG): Project No. 386119226. V.~B.~B. is partially supported by the Conselho Nacional de Desenvolvimento Científico e Tecnológico (CNPq)-Brazil, through the Research Project No. 307211/2020-7. K.~D.~K acknowledges support from the Tübingen-Nottingham Joint Seedcorn Fund.

\bibliography{literature}

\end{document}